# One-step deposition of nano-to-micron-scalable, high-quality digital image correlation patterns for high-strain in-situ multi-microscopy testing

**J.P.M. Hoefnagels\*, M.P.F.H.L. van Maris, T. Vermeij**

*Eindhoven University of Technology, the Netherlands*

\* *j.p.m.hoefnagels@tue.nl*

## Abstract

Digital Image Correlation (DIC) is of vital importance in the field of experimental mechanics, yet, producing suitable DIC patterns for demanding *in-situ* (micro-)mechanical tests remains challenging, especially for ultra-fine patterns, despite the large number of patterning techniques reported in the literature. Therefore, we propose a simple, flexible, one-step technique (only requiring a conventional physical vapor deposition machine) to obtain scalable, high-quality, robust DIC patterns, suitable for a range of microscopic techniques, by deposition of a low melting temperature solder alloy in so-called 'island growth' mode, without elevating the substrate temperature. Proof of principle is shown by (near- )room-temperature deposition of InSn patterns, yielding highly dense, homogeneous DIC patterns over large areas with a feature size that can be tuned from as small as ~10 nm to ~2 µm and with control over the feature shape and density by changing the deposition parameters. Pattern optimization, in terms of feature size, density, and contrast, is demonstrated for imaging with atomic force microscopy, scanning electron microscopy, optical profilometry and optical microscopy. Moreover, the performance of the InSn DIC patterns and their robustness to large deformations is validated in two challenging case studies of *in-situ* micro-mechanical testing: (i) self-adaptive isogeometric digital height correlation of optical surface height profiles of a coarse, bimodal InSn pattern providing microscopic three-dimensional deformation fields (illustrated for delamination of aluminum stretchable interconnects on a polyimide substrate) and (ii) DIC on scanning electron microscopy images of a much finer InSn pattern allowing quantification of high strains near fracture locations (illustrated for rupture of a polycrystalline Fe foil). As such, the high controllability, performance and scalability of the DIC patterns, created by island growth of a solder alloy, offers a promising step towards more routine DIC-based *in-situ* micro-mechanical testing.

## DIC Keywords

Digital Image Correlation, *in-situ* micromechanical testing, scanning electron microscopy, optical profilometry, experimental mechanics, DIC pattern

# Introduction

Digital Image Correlation (DIC) is an essential technique in today's field of experimental mechanics. It allows full-field measurement of displacements, from which, e.g., strains can be derived, with high (sub-pixel) accuracy, and can be applied on a wide range of tests without requiring direct (mechanical) contact, while being highly robust to noise [1-3]. DIC has proven to be a broad multi-purpose method, being employed in a wide range of applications and investigations, ranging from analysis of objects as large as airplanes and bridges [4,5], to detailed identification of crystal plasticity and damage mechanisms on the micro or even nanoscale [6,7]. Moreover, DIC is extremely versatile, as it can be performed on various types of images, acquired with regular camera, optical microscope [e.g., 8,9], Scanning Electron Microscope (SEM) [e.g. 10], or transmission electron microscopy [11]. Yet, DIC is not limited to (2D) in-plane displacement measurements but can readily be extended to measure the three-dimensional displacement field of a surface, with (multi-camera) stereo-DIC [e.g. 12] or Digital Height Correlation (DHC), where the surface height profiles can be acquired with an optical profilometer [e.g. 6, 13] or an Atomic Force Microscope (AFM) [e.g., 14,15]. DIC can even be applied to measure the 3D displacement field of a volume, e.g., by correlating X-ray computed tomography images [e.g. 16]. The correlation on the acquired images or height data is traditionally performed through the correlation of local regions (subsets or windows), sometimes called 'local DIC'. In contrast, in 'global DIC' the correlation is performed over the complete Field Of View (FOV), while typically material continuity is assumed over the complete FOV [17]. Finally, these full-field measurements can directly be coupled in various ways to numerical simulations or theoretical models for the identification of (material) parameters [e.g. 18-21].

A crucial requisite for proper DIC analysis is a high-quality pattern that provides enough contrast at the scale of interest (locally and/or globally) for the DIC algorithm to be able to track. While the natural features of the sample are satisfactory in some cases, e.g. for multi-phase materials [22-24], their change during the experiment is often detrimental to the quality of the correlation [25]. Therefore, in most cases, high quality DIC analysis requires application of a pattern on the sample surface. For optimal correlation accuracy, the pattern should consist of a homogeneous, high-density distribution of features (speckles) with the size of a few pixels in order to achieve a small pattern correlation length and high mean intensity gradient [26, 27]. Therefore, the patterns must be tailored to the specific requirements of the experiment, such as the type of microscopy (or imaging) technique, the length scale of observation (both the smallest scale and the total FOV) at which the correlation should be performed, the ability to sustain high (localized) strains without showing pattern degradation or changes, etc. Application of such a high-quality pattern is typically far from trivial, often limiting the spatial resolution and accuracy of the displacement and strain field, especially for the nanoscale patterns needed for high-magnification experiments. Moreover, the pattern must be applied without altering (the microstructure and deformation of) the underlying specimen in any way or form. For example, a heat treatment can easily cause undetected re-arrangement of dislocations in a (poly-)crystal thereby affecting the micro-mechanical behavior, which obviously should be avoided.

For DIC of images obtained with a regular camera (or low-magnification optical microscope), simple techniques such as spray painting or using an airbrush typically provide a good pattern [28], yet, correlations at higher magnification (high-magnification optical microscopy, optical profilometry, SEM or AFM) place more stringent requirements on the DIC pattern. In the literature, a large number of techniques exist to apply micro- or nanoscale DIC patterns [29-40]. Two types of nanoscale patterning techniques stand out. First, nano-particles (of different materials and sizes) can be dispersed over the sample surface. This can be achieved, e.g., by so-called 'drop casting' or 'mistification of droplets' of nano-particles in a solvent, however, a homogeneous particle density without clusters is notoriously difficult to achieve due to Marangoni flow in the droplets while the solvent evaporates at the surface [32]. Therefore, alternative particle dispersion techniques have been proposed, e.g. (wet) self-assembly of gold particles [33, 34] or pressing of colloidal silica particles into the surface during chemically assisted polishing [35]. While these alternative particle dispersion techniques have clearly proven their merits, they are still sensitive to clustering of the particles, which can only be completely avoided at the expense of a lower feature density. Moreover, they rely on (lengthy or cumbersome) processing steps that are sensitive to operator input, therefore, good reproducibility is not readily achieved. Second, nanoscale pattern application by deposition of a metal (e.g., gold or silver) has been proposed, where first one or more layers of homogeneous thickness are deposited followed by a heat treatment (at high temperature and/or assisted by a chemical reaction) to instigate remodeling of the layer into separate islands [29, 36, 37, 40]. The importance of all of these patterning techniques to the community is undisputed, as can be clearly seen from the high number of citations to most of the above-mentioned references. Yet, truly ultra-fine patterns (with a feature size around 10 nm) are not readily achieved. Moreover, even the best patterns typically still contain significant pattern-free area between the particles or islands, often resulting in nearly black & white images with few intermediate grey values. Moreover, most of these techniques have been optimized for one specific microscopy technique, length scale or sample geometry, without demonstrated flexibility for subtle modifications in size, size-range, density, etc. Finally, as explained above, heat treatments, pressure, or chemical reactions are best avoided. Considering these concerns, there is still a clear interest for an easy-to-implement, (near-)room-temperature, one-step patterning method that can produce high-quality micro- to nanoscale DIC patterns with high reproducibility. It is important that the processing method provides sufficient freedom to allow uncomplicated tuning of the shape and size of the pattern features, in order to adhere to a wide range of DIC environments and microscopic techniques, without being constrained to low strains.

In this work, we explore the suitability of one-step deposition of metal islands (using Physical Vapor Deposition, PVD) in the so-called 'island growth' (Volmer–Weber) deposition mode, as opposed to first depositing a homogeneous layer that subsequently requires thermal/chemical processing. Island growth requires a large surface diffusion length of the metal adatoms that typically only occurs at a very high substrate temperature [41-43]. Here, however, we set the requirement that the pattern must be applied through a single short-duration deposition step without requiring substrate temperature elevation (also because most commercial deposition systems come without substrate temperature control). Without specimen heating, direct island growth may still be achieved by selecting (i) metal adatoms with a high surface mobility (e.g. (eutectic) low-melting temperature (solder) alloys) in combination with (ii) a plasma that boosts the adatom surface diffusion length as each adatoms receives the kinetic energy from multiple ions. By sufficiently increasing adatom surface diffusion length the growth of islands is activated from the start of the deposition. An added advantage of island growth would be that, for very short deposition times, the islands should be very small and nearly touching each other, whereas by increasing the deposition time the size of the islands can be tuned to the DIC application at hand. In this work, a simple DC magnetron sputter deposition machine is employed to demonstrate that high quality patterns are easily accessible in a reproducible manner, while it is shown that good patterns can be achieved for a wide range of deposition/plasma settings. This makes the exact specifications of the deposition machine unimportant.

## Patterning Method and Results

A Torr CRC-600 system without substrate temperature control is used to perform planar direct current (DC) magnetron sputter deposition. This is a PVD method in which, under vacuum, (clusters of) atoms are ejected from a target, by positively charged Ar ions, and accelerated towards the sample surface, where they collect to ultimately form a layer, see Figure 1. We believe, however, that any sputter deposition machine could have been used. As explained in the Introduction, the feature (island) size is largely governed by the mobility of the metal adatoms on the surface, i.e. the adatom surface diffusion length, to form large islands. According to [43], the adatom mobility is increased when a low pressure is administered, since the target atoms retain more kinetic energy during the transport from target to substrate due to fewer collisions with the background gas. Also, high temperatures of the adatoms with respect to the melting temperature of the target material is reported to increase the adatom mobility and therefore to provoke island growth [43]. Here, however, the aim is to avoid substrate temperature elevation, therefore instead, a solder alloy with (very) low melting temperature is selected. Accordingly, two deposition materials are explored in this study: (i) a simple solder alloy (SAC305, Sn96.5Ag3Cu0.5, melting temperature 217 °C) and (ii) an InSn alloy (In52Sn48, melting temperature 118 °C), which are both melted in a crucible for insertion as a target in the deposition system. The morphology of the DIC pattern features can be manipulated by varying several processing parameters of the deposition process:

the chamber pressure, the plasma current, and the sputter time. Finally, although the used deposition system comes without a substrate heater, the effect of a mild substrate temperature elevation on the island morphology is still explored, for some of the patterns, by inserting a heated block as substrate holder.

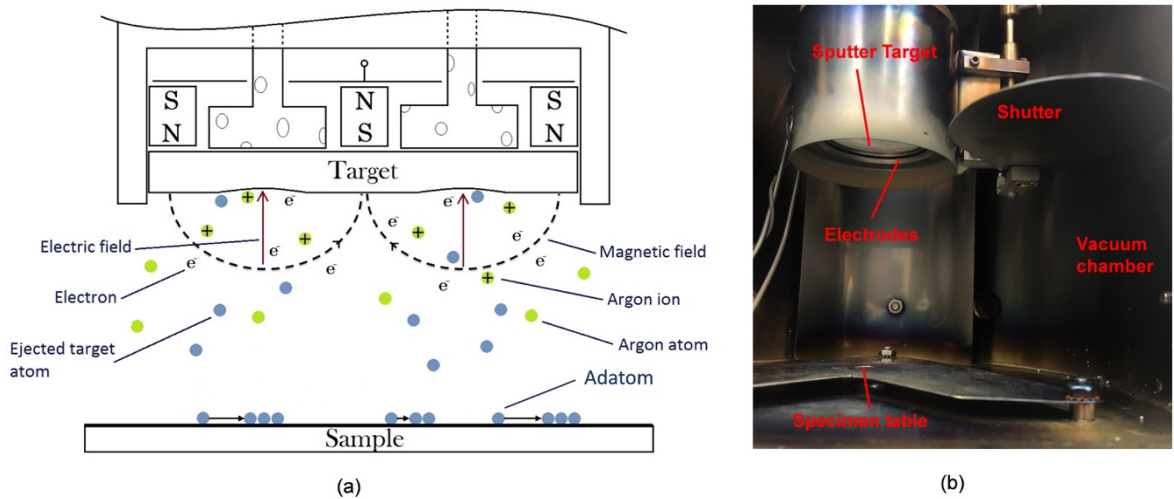

*Figure 1: (a) schematic illustration of the DC magnetron sputter deposition process, (b) image from inside the deposition chamber with specifications of relevant components.*

First, we try to achieve a DIC pattern using the SAC alloy on a Si wafer, by performing a deposition at a chamber pressure of 100 mTorr and a sputter current of 20 mA for 3 minutes, with the substrate at room temperature. For this first deposition try no surface features could be observed in the SEM. Therefore, the sample was shortly heated to 300 °C, with the final result shown in Figure 2. A pattern with ultra-fine nano-scale features is observed with a density and contrast that appears to be challenging those reported in the literature, thereby making this pattern certainly suitable for *in-situ* SEM testing at high magnification for sub-micron DIC resolution. However, the clusters, which probably have been extracted directly from the target, and the dark residue spots, which likely were induced by oxidation during the heat treatment, show that the pattern is not perfect. Most importantly, the post-deposition heat treatment, which is required to remodel the homogeneous SAC alloy layer into islands, should be avoided as explained in the Introduction. Therefore, further attempts with the SAC alloy were abandoned.

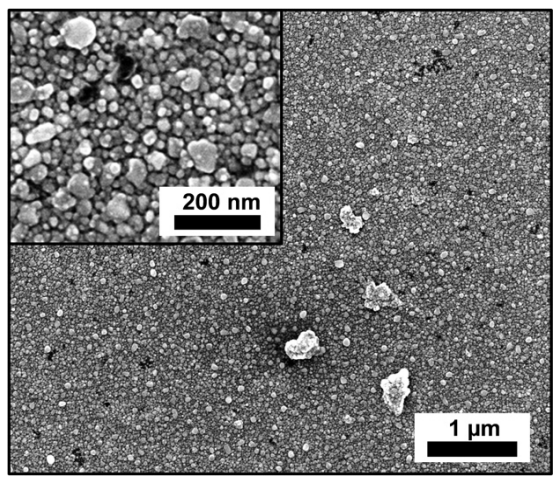

*Figure 2: SAC pattern, with post-deposition heat treatment to 300 °C, imaged in secondary electron (SE) mode in a FEI Quanta 600 SEM, showing a highly dense nano-scale pattern, but with a significant amount of large clusters and dark residue spots.*

Next, the InSn alloy was tested, as it has an even lower melting temperature of only 118 °C and is, therefore, expected to provide a much larger adatom surface diffusion length. For each of the wide range of deposition settings that were initially explored with the substrate at room temperature, a fine scale pattern of islands was deposited. This shows that the concept of direct island growth of a low-melting-temperature metal is indeed feasible and that this concept is not limited to a narrow range of deposition parameters. Subsequently, a representative range of InSn DIC patterns has been deposited using the six different sets of deposition parameters that are provided in Table 1. Each resulting pattern is imaged in an SEM at high resolution, as shown in Figure 3. As can be seen, all deposition settings, i.e. the substrate material (Si wafer and polyimide (PI); and iron in the second test case below), sputter current, chamber pressure, sputter time, and substrate temperature, were varied, confirming that high quality patterns can be achieved for a wide range of deposition/plasma settings (and most probably different types of deposition machines). From these SEM images, the feature size range is qualitatively estimated and included in Table 1. The table also shows the correlation length, which is calculated as the standard deviation of the Gaussian function fitted to the peak in the autocorrelation function of the pattern. Each pattern is homogeneous over a very large area. Moreover, each pattern has a distinct length scale, as seen from both the feature size range and correlation length. The feature sizes ranges between ~10 nm and ~2 μm, while the correlation length varies between 6 nm and 159 nm. The difference between feature size and correlation length can be explained by the fact that the correlation length is sensitive to the sharpness of the edges of the features, which are predominantly smaller than the feature size. The usefulness of these patterns for performing high quality DIC will be shown below by testing pattern **c** and **e** in *in-situ* DIC tests using, respectively, SEM and optical profilometry.

Generally, the following effects of the deposition parameters were observed: higher sputter currents and deposition times yield larger features while a higher chamber pressure results in smaller features. Note that, comparing pattern **a** to **b**, the sputter current and chamber pressure are both significantly increased, while the size of the features are similar, whereas one can clearly distinguish a difference in morphology between them, with, e.g., sharper edges in pattern **b** than in **a**. This shows that two deposition parameters can be simultaneously varied to control a single property of the DIC pattern. It also indicates that the patterning method can indeed be well tuned to a demanding imaging technique which might require a specific type of feature size, shape or contrast. Additionally, when the features become larger, there seems to be a natural development towards a bimodal distribution of island sizes, which can be observed in pattern **c** and **d**, but is especially clear in pattern **e** and **f**, with ratios between feature sizes of up to 20. Such a bimodal DIC pattern can have large benefits when deformation fields need to be measured at multiple scales, by applying DIC on images using the same pattern but captured at multiple magnifications [44]. Note that these bimodal patterns appear as highly dense, homogeneous patterns when imaged at the lower magnification, as shown in the examples of Figures 5 and 6 below. Therefore, the bimodal character is not detrimental to the pattern quality when imaged at a single magnification that is appropriately selected.

*Table 1: Deposition parameters and pattern properties for the range of six InSn DIC patterns. For the wide range of deposition settings that were probed, including the six settings in this Table, a homogeneous nano-sized pattern is consistently achieved.*

| sample | substrate material | current [mA] | pressure [mTorr] | time [min] | T [°C] | feature size range [nm][1] | correlation length [nm][2] |
|---|---|---|---|---|---|---|---|
| **a** | Si | 10 | 8 | 5 | 20 | ~10 – ~30 | 6 |
| **b** | Si | 40 | 100 | 15 | 20 | ~15 – ~60 | 8 |
| **c** | Si | 15 | 8 | 3 | 20 | ~20 – ~100 | 24 |
| **d** | Si | 15 | 8 | 5 | 20 | ~30 – ~200 | 33 |
| **e** | PI | 20 & 60 | 8 | 2 & 7 | 80 | ~100 – ~1000 | 62 |
| **f** | PI | 50 | 8 | 15 | 20 | ~100 – ~2000 | 159 |

[1] *qualitative estimation from SEM images in Figure 3*
[2] *standard deviation of the Gaussian function fitted to the peak of the autocorrelation function of the patterns from Figure 3*

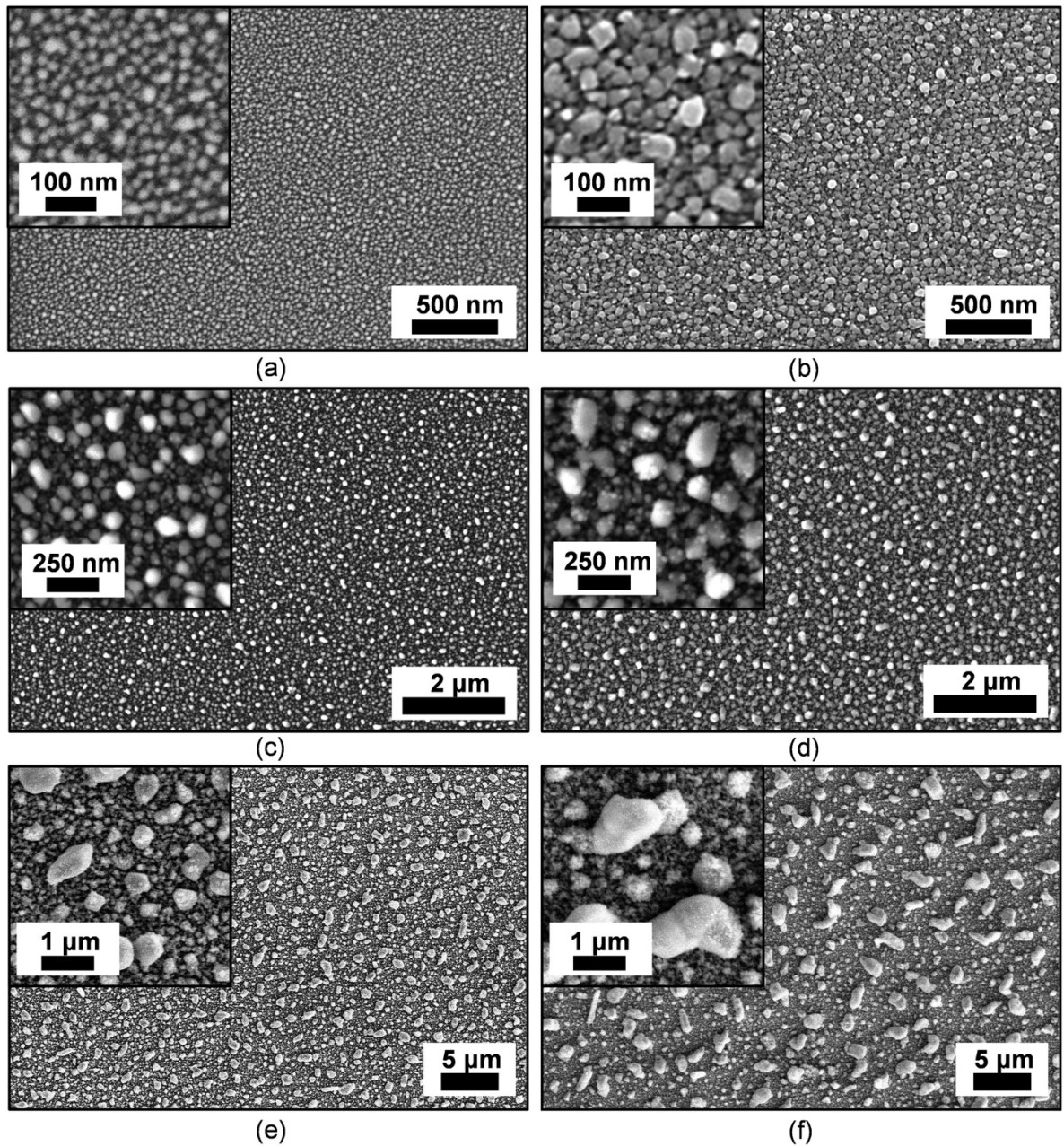

*Figure 3: Secondary Electron SEM scans of 6 different InSn DIC patterns with increasing feature size, produced with the deposition parameters listed in Table 1. Image a,c and d are captured with a Tescan Mira 3; image b, e and f are captured using a FEI Quanta 600.*

    The small features in pattern **a** suggest that this pattern could also be well suited for DIC on AFM height profiles, in order to conduct DIC with higher spatial resolution than feasible with SEM images, but this would require the pattern to have a sufficient and clear height contrast. To this end, Figure 4 shows an AFM scan (captured with a Digital Instruments DI 3100) of pattern **a**, clearly confirming the high, homogeneous feature density of this pattern. The features have round shapes, as expected for an 'island growth' deposition pattern, with height variations in the range of a few nanometers. Moreover, the side slopes of the features are not too steep, thereby enabling accurate AFM imaging. The height contrast that is observed indeed suggests that pattern **a** is not only suited for *in-situ* SEM testing, but also provides good height contrast for nanoscale DIC using an AFM. This allows, e.g., the combined imaging with AFM and SEM during an *in-situ* mechanical test, which has not yet been demonstrated in the literature.

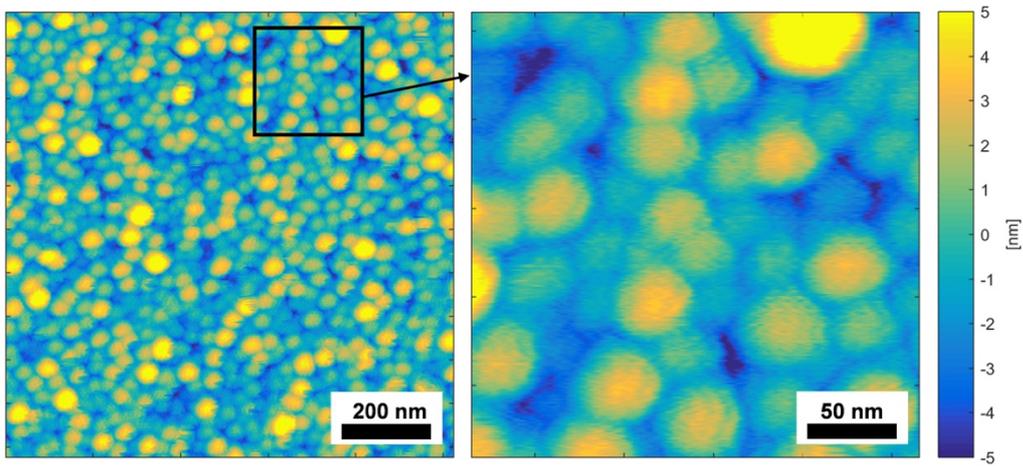

*Figure 4: Nano-scale AFM scan of InSn DIC pattern **a**, from Figure 3a and table 1.*

With the increase in feature size from pattern **a** to **f**, the slope of the side edges of the larger islands inevitably also becomes steeper, which could become a problem for optical profilometry as it needs a courser pattern than AFM. Therefore, with pattern **e** an example is shown of a deposition performed at an elevated substrate temperature of 80 °C. As expected, the pattern size remains (visibly) unchanged as a result of this relatively small temperature increase, as this is insufficient to significantly increase the adatom surface diffusion length. However, it was found that this mild substrate temperature increase is sufficient to slightly round off the edges of the islands and thus to decrease the side slope angle. This circumvents the need to eliminate incorrect measurements points (NaN's) from the optical surface height profiles, as will be shown in Figure 6 below.

Alternatively, the patterns should also be applicable for DIC in combination with a less complex imaging technique, such as optical microscopy. Hence, we briefly investigated how well, e.g., pattern **e** is suitable for this purpose. To this end, Figure 5 shows Nomarski Interference Contrast (NIC) mode images of pattern **e** at two magnifications (captured with a Zeiss Axioplan 2). Using a 20X objective in Figure 5a, pattern **e** shows a highly dense and homogeneous pattern over a large area. Moreover, the inset shows that sufficient contrast is retained for DIC on this scale. Additionally, using 5 times higher magnification in Figure 5b on the same location, smaller details can still be observed with similar quality to the SEM image in Figure 3e, owing to the high sensitivity of NIC to surface roughness. This example shows that, with a rather simple optical microscope, a single pattern contains sufficient details from sub-micrometer to millimeter range, confirming the powerful ability of this patterning technique for *in-situ* multi-scale DIC experiments. Similarly, pattern **f** could be used for two-scale imaging with two regular optical cameras.

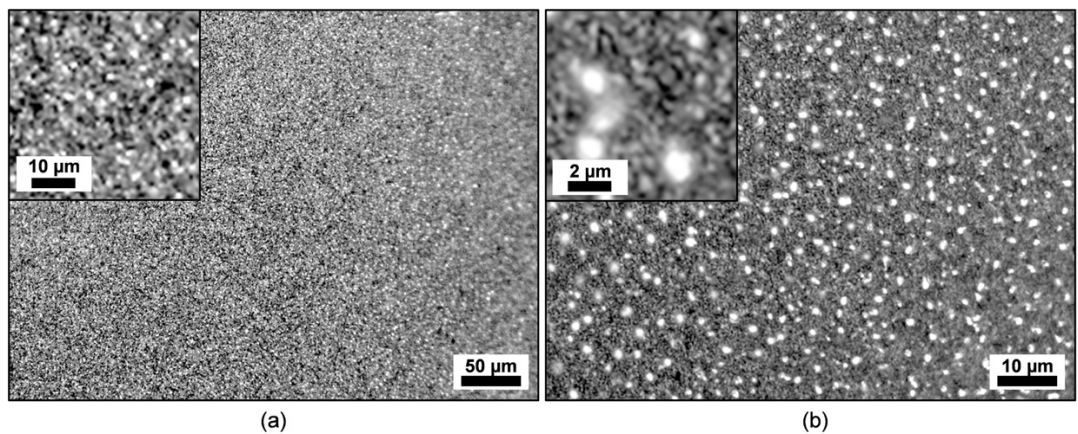

*Figure 5: Optical Microscopy images of InSn DIC pattern **e**. High contrast on multiple scales is shown with (a) a 20X objective NIC image, showing a homogeneous pattern over a large area and (b) a 100X objective NIC image in the same location, in which sub-micron sized features can be distinguished. Note that the blurring on the right side of the images is caused by the out-of-plane curvature of PI foil, which is the substrate of this pattern.*

# Proof of principle: *in-situ* optical profilometry and SEM testing

While all the patterns are expected to perform well during DIC experiments, we demonstrate this with two challenging *in-situ* micro-mechanical tests analyzed with appropriately chosen DIC methods: (I) quantification of the three-dimensional surface displacement field during delamination of a thin aluminum interconnect adhered to a polyimide substrate and (II) analysis of micron-scale plasticity, damage localization, and fracture mechanisms in a polycrystalline iron foil.

For case study I, we are interested in the failure limits of stretchable electronics, and specifically, how the delamination between aluminum and polyimide occurs during, e.g., a global tensile test. As delamination in this test predominantly yields out-of-plane displacements, stereo-DIC could be applied, to measure the three-dimensional surface displacement field (i.e. the combined in-plane and out-of-plane displacement fields) on the specimen [45, 46]. However, for stereo-DIC applied to images captured with an optical stereo microscope, the out-of-plane displacement resolution is limited due to the small viewing angle between the two cameras. Here, however, Digital Height Correlation (DHC) is applied to surface height profiles measured with an optical profilometer [6, 13, 47-49]. With this technique 3D surface displacement fields can be measured with very high out-of-plane displacement resolution (down to a few nanometer [13]), as it uses high-resolution quantitative height data instead of qualitative grey-scale values in the correlation algorithm. The height profile of an as-received sample is shown in Figure 6a, which clearly shows insufficient local contrast for a proper DHC correlation (which is confirmed by preliminary tests, not shown here), while the steep edges also show erroneous height measurements, which would be problematic for DHC. Applying InSn pattern **e** results in the topography in Figure 6b, where it is seen that the side edges of the deposited InSn islands and the aluminum interconnect structure are steep enough for a good DHC pattern but still smooth enough such that erroneous measurement points are prevented. An overview of the geometry of the aluminum interconnect on the polyimide substrate, the global tensile direction, and the considered region of interest is given in Figure 6c. Tensile experiments on the stretchable electronics samples were performed *in situ*, using a Kammrath & Weiss micro-tensile stage under a Sensofar Plµ 2300 optical profilometer in confocal mode, with a 150X objective to obtain the highest possible magnification. The tensile test was conducted up to 40% applied global strain. For the stretchable interconnect structures, limited *a-priori* knowledge regarding the three-dimensional deformation of the samples upon stretching, especially the moment of buckling, was available. Therefore, 'self-adaptive isogeometric DHC', as developed in [49,50], is employed, i.e. isogeometric shape functions are used for the representation of the 3D displacement fields (in a 'global DIC' setting) in combination with a self-adaptive refinement procedure based on hierarchical refinement of the shape functions.[1] At the start of the deformation, a simple NURBS mesh is applied separately to the interconnect and the surrounding substrate on both sides, as shown in Figure 6d. During intermediate deformation steps (not shown here), the NURBS mesh is successfully refined in an automated fashion due to the localized behavior, showing good convergence towards a low height residual (not shown). Figure 6f shows the final deformation step with a locally double-refined mesh drawn in the deformed configuration. The corresponding *x*-, *y*-, and *z*-displacement fields are given in Figure 6e,g,h, in the reference configuration, which shows an intricate buckling pattern. This three-dimensional displacement field of the buckling pattern is quantitatively compared to multi-scale FEM simulations including Cohesive Zone interface elements to capture the delamination behavior in Ref. [52]. There it was found that this measured buckling pattern can be reproduced numerically by employing a higher fracture toughness in shear opening than in normal opening.

In case study II, we investigate the evolution of micro-plasticity and damage during deformation of a polycrystalline Fe foil, by performing an *in-situ* tensile test, using a Kammrath & Weiss micro-tensile stage, inside a Tescan Mira 3 SEM, employing in-beam SE imaging. The Fe foil was electro-polished in advance, to expose the microstructure. Subsequently, an InSn pattern was deposited on the surface, utilizing the sputter deposition parameters according to pattern **c** from table 1. SEM scans of 3096x3096 pixels were taken before and during deformation (corresponding to a global strain of 1%, 2.5% 3.5% and 5%) over a 50x50 μm² FOV, resulting in a pixel size of ~16 nm. Figures 7a and 7b, respectively, show the SEM images without deformation and at the last deformation step. We focus on one grain with a particular surface topology, which deforms the most during the experiment (as shown by the insets). In this grain a high-quality DIC pattern is observed that nicely follows the locally steep surface topology. Importantly, the grain boundaries and triple junctions can still be seen through the pattern. This allows to easily find back the designated analysis region that was characterized before pattern application by, e.g., Electron BackScatter Diffraction (EBSD) (not performed here). Typically, during *in-situ* SEM testing, the contrast and brightness settings of the SE detector are adapted to counteract the reduction in contrast and brightness due to continuous electron beam-induced carbon deposition, also called carbon contamination. Here, however, the contrast and brightness settings are kept deliberately unchanged during the whole experiment to demonstrate the resilience of the DIC pattern against carbon contamination. For this case study including fracture, a 'local', subset-based DIC analysis is the preferred choice, as material continuity is not assumed. A commercial DIC software package (VIC-2D from Correlated Solutions) was used to perform subset-based DIC on the SEM images, using a subset size and step size of, respectively, 21 and 1 pixel. Green Lagrange strains are computed through nearest-neighbour differentiation without filtering, essentially corresponding to a strain window of 1 data point. These settings correspond to a small 'Virtual Strain Gauge' (VSG) size of ~336nm when using the formula of $L_{VSG} = (L_{window} - 1)L_{step} + L_{subset} = 21$ pixels proposed in [53]. This results in the equivalent strain fields shown in Figure 7c,d,e,f. Localized deformation bands can be discerned with great detail, without any degradation of the pattern at 1% global strain. During further deformation, localization continues, and the heterogeneity of the strain field increases significantly. Physical cracks appear in the foil, however, the correlation remains successful directly at the edges of the cracks, where strains of over 100% are measured. The occurrence of cracks has been confirmed by post-mortem SEM observation of the back-side of the foil (not shown). The stability of the DIC pattern between fractured parts of the material can be of crucial importance when, e.g., propagation of damage is studied, which is not trivial [25]. In the literature, a VSG size between 100 and 200 pixels (with an optimum at 141 pixels) is recommended as a compromise between good image correlation stability for large VSG and good spatial resolution (with low underestimation of the maximum strain) for a small VSG [54]. With our pattern, however, high image correlation stability is achieved for a much smaller VSG (of 21 pixel). This provides the possibility to measure strains at a much higher spatial resolution in terms of pixels, while the images have been captured with a relatively small pixel size of ~16nm (compared to most *in-situ* SEM testing publications).

---

[1] In this procedure, Non-Uniform Rational B-Splines (NURBS), which can represent various geometrical shapes, are used for the shape functions and geometrical parametrization, offering high computational efficiency, accuracy and robustness [51]. Upon localized deformation, the hierarchical refinement of the shape functions locally increases the number of degrees of freedom in an objective manner, by evaluation of the height residual obtained in the previous iteration.

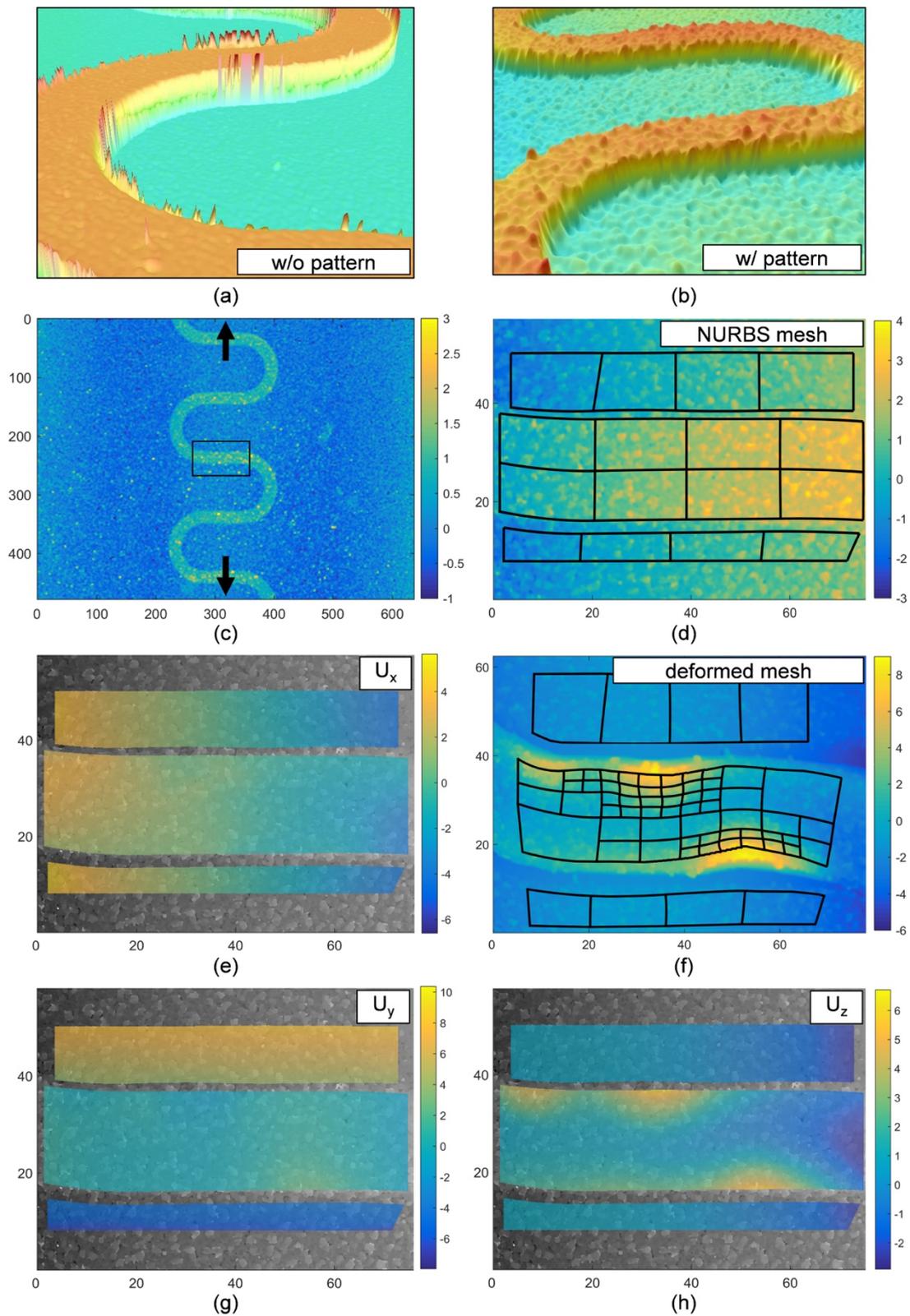

*Figure 6: Characterization of microscale deformation of stretchable electronics using self-adaptive isogeometric DHC of in-situ optical profilometry height data: 3D View of a topography of a meandering aluminum interconnect adhered to a polyimide foil substrate, (a) without and (b) with an applied InSn DIC pattern (pattern **e** shown in Figure 3e). (c) Overview of the topography and testing condition (also showing the global tensile direction and the region of interest) of the interconnect. (d,f) Graphical representation of the NURBS mesh used for DHC, (d) the initial mesh drawn over the initial surface topography and (f) the self-refined mesh after convergence of the adaptive isogeometric DHC drawn in the deformed configuration over the final topography. (e,g,h) The successfully correlated three-dimensional displacement field, plotted in the reference configuration, consisting of the displacement field in (e) x, (g) y, and (h) z (out-of-plane) direction. All dimensions are in µm.*

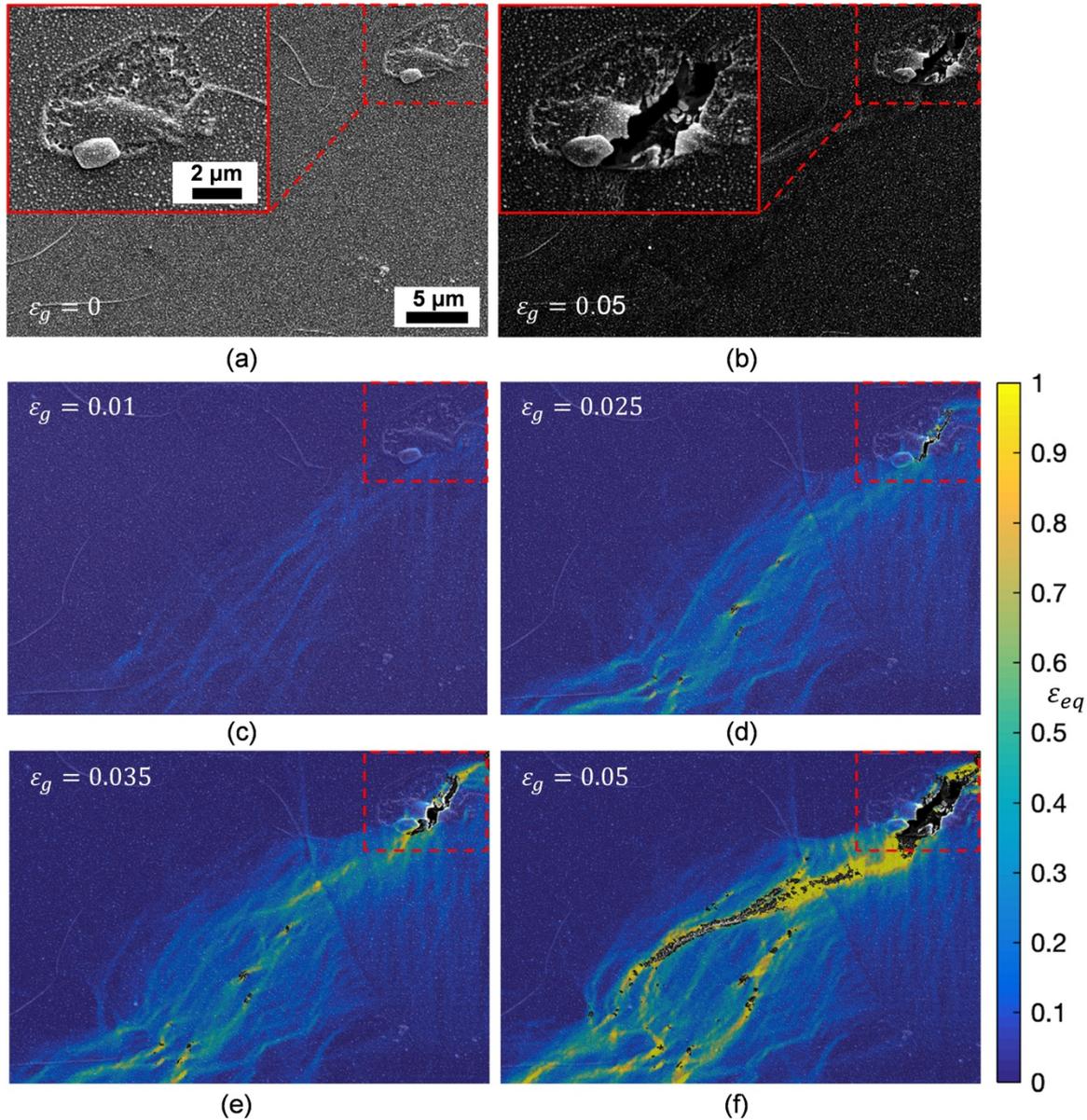

*Figure 7: Overview of an in-situ SEM tensile test, processed by subset-based local DIC, on pure Fe, using in-beam SE scans. (a) Reference SEM scan at the start of the in-situ test, showing the InSn DIC pattern (created with the processing parameters of pattern **c** from table 1), directly deposited on the electro-polished surface of polycrystalline Fe, while the inset shows one particular grain. (b) The same area after uniaxial tension up to 5% global strain (with a large crack in the grain in the inset), subject to major loss of contrast due to contamination (contrast/brightness of the in-beam SE detector was not changed during the test). (c,d,e,f) The equivalent strain fields at, respectively, 1%, 2.5%, 3.5% and 5% global strain, as computed by subset-based local DIC.*

# Discussion: applicability of the patterning method

For a DIC patterning technique, the suitability to various substrate materials is an important concern. In table 1 and figure 3 we show that Silicon and Polyimide can be used, which proves that a conductive substrate material is not a requirement. Alternatively, case study I and II demonstrate that aluminum and iron are also suitable as a substrate. As with all forms of low-pressure deposition, the InSn adatoms will have a finite (i.e. non-zero) sticking probability on all surfaces, therefore a pattern should form on all materials. The only real limitation is that the specimen does not damage or degrade in the environment of the sputter deposition process, i.e. low pressure, electron charging, and mild ion bombartment, which limits the applicability to delicate and moisture-sensitive specimens such as biological samples, hydrogels, etc. Moreover, it should be noted that the adatom surface diffusion length will not be equal for different materials, hence, the same deposition parameters cannot guarantee the exact same morphology and size of the InSn pattern for different substrate materials (even though pattern **c** appears similar when deposited on silicon and iron). Therefore, for multi-material samples, such as e.g. micro-electronical components, it could potentially be more challenging to attain a highly homogeneous pattern, locally, over the different materials. Note that a continuous thin film of, e.g., Titanium [55] could be added underneath the InSn pattern to alleviate this concern, although care should be taken that such a continuous film can sustain the deformation of the underlying specimen inside localization or damage zones.

We have not observed any direct effect of SEM imaging on the DIC pattern in terms of charging or local heating. For a continuous thin film, the electron beam energy may induce remodeling into separate islands to reduce the total energy of the system (similar to the concept of Au or Ag remodeling as discussed in the Introduction), however, because the InSn is directly deposited in the low-energy configuration of separate islands, it is not surprising that this pattern is stable under electron beam imaging. Probably, however, the InSn pattern is not stable above its (low) melting temperature, thus restricting testing at higher temperature. There may be two options to circumvent this issue: (i) a nanometer thin layer of a material with higher melting temperature can be deposited on top of the InSn pattern, such that this thin film takes the shape of the underlying pattern and retains this during high temperature testing, or (ii) a sputtering material with a melting temperature above the testing temperature can be chosen, while assuring the island growth sputter deposition mode by performing the deposition at a temperature close to the melting temperature.

The InSn alloy shows great promise as room-temperature sputtering material, yielding features from nanometer to micrometer size in a controllable manner. Nonetheless there should be good alternatives for room temperature deposition and testing, e.g., other low melting temperature (solder) alloys such as 49Bi21In18Pb12Sn (melting temperature of 58°C). Finally, after the *in-situ* test has been finished, a detailed analysis of the substrate surface underneath the pattern is often of interest. To this end, a dedicated (electro-)chemical etching step for InSn, e.g., the one suggested in Ref. [56], may be employed to completely remove the pattern to enable, e.g., post-mortem BSE and/or EBSD analysis.

## Conclusions

In this work, we demonstrate the feasibility of a relatively simple technique to obtain a, controllable, high quality DIC pattern for multi-microscopy testing, over a range of scales, on different substrates (Si, polyimide, aluminum, iron). Using a low melting temperature (InSn) solder alloy, we provide the ability to create dense, high-quality DIC patterns over large areas with feature sizes ranging from ~10 nm to ~2 µm, without resorting to elevated substrate temperatures, challenging the most sophisticated patterning techniques available in the literature. The capability of controlling several properties of the pattern, such as feature size, size range and shape, is achieved by changing two key sputter deposition parameters to fine-tune the properties of the patterns. All DIC patterns considered here were imaged using scanning electron microscopy, showing good contrast in combination with a high, homogeneous feature density. Moreover, the finest pattern seems to have an optimal morphology for nano-scale DIC on AFM height profiles, whereas the patterns with larger features show good potential for multi-scale DIC using a regular multiple-magnification optical microscope (or multiple optical cameras).

Finally, the robustness and applicability of the DIC patterns was tested on two challenging case studies. (i) A relatively new type of DIC method, i.e. self-adaptive isogeometric digital height correlation on optical surface height profiles, was investigated. For this method, which requires a very specific type of pattern (micrometer sized speckles, good height contrast, and smooth slopes at the feature edges), it was demonstrated that a dedicated pattern could be deposited, yielding high quality three-dimensional surface deformation fields. (ii) At the nano-scale, it can be challenging to elucidate complex micro-plasticity, damage and fracture mechanisms during *in-situ* experiments, due to conflicting requirements of high spatial resolution of the strain field and high robustness of the image correlation at high strain.

It was demonstrated, for an *in-situ* SEM tensile test on a polycrystalline pure Fe foil, that the InSn DIC pattern with nanoscale features is ideal for sustaining and quantifying high local strains near fractured areas.

In summary, the high controllability and performance of DIC patterns created by physical vapor deposition of a low melting temperature solder alloy, which only requires a single deposition step with a conventional sputter deposition system, provides a promising step towards more routine *in-situ* DIC experiments that normally require highly optimized and complex patterning methods.

# Acknowledgements

The authors greatly acknowledge Rob Fleerakkers, Chaowei Du, Sandra van de Looij-Kleinendorst, Uriel Hoeberichts, Jan Neggers, Philip Reu, Benoît Blaysat, Salman Shafqat and Samaneh Isavand for their contributions and/or discussions. Part of this research was carried out under project number S17012b in the framework of the Partnership Program of the Materials innovation institute M2i (www.m2i.nl) and the Technology Foundation TTW (www.stw.nl), which is part of the Netherlands Organization for Scientific Research (www.nwo.nl).